\documentclass[prepare]{elsarticle}
\usepackage{color}

\journal{Journal of \LaTeX\ Templates}

\begin{document}

\begin{frontmatter}

\title{\boldmath R \& D of prototype iTOF-MRPC at CEE}

\author[address1,address2]{Y.Zhou}

\author[address1,address2]{D.Hu\corref{correspondingauthor}}
\ead{hurongr@ustc.edu.cn}

\author[address1,address2]{X.Wang}

\author[address1,address2]{M.Shao\corref{correspondingauthor}}
\ead{swing@ustc.edu.cn}

\author[address1,address2]{L.Zhao}

\author[address1,address2]{Y.Sun}

\author[address1,address2]{J.Lu}

\author[address1,address2]{H.Xu}

\address[address1]{State Key Laboratory of Particle Detection and Electronics, University of Science and Technology of China, 96 Jinzhai Road, Hefei 230026, China}
\address[address2]{Department of Modern Physics, University of Science and Technology of China (USTC), 96 Jinzhai Road, Hefei 230026, China}
\cortext[correspondingauthor]{Corresponding author}


\begin{abstract}
The cooling storage ring (CSR) external-target experiment (CEE) is a spectrometer running at the Heavy Ion Research Facility (HIRFL) at Lanzhou. The CEE is the first large-scale nuclear physics experimental device by China to operate in the fixed-target mode with an energy of $\sim$ 1 GeV. The purpose of the CEE is to study the properties of dense nuclear matter. CEE uses a multi-gap resistive plate chamber (MRPC) as its internal time-of-flight (iTOF) detector for the identification of final-state particles. An iTOF-MRPC prototype with 24 gaps was designed to meet the requirements of CEE, and the readout electronics of the prototype use the FPGA-based time digitization technology. Using cosmic ray tests, the time resolution of the iTOF prototype was found to be approximately 30 ps. In order to further understand how to improve the time resolution of MRPC, ANSYS HFSS was used to simulate the signal transmission process in MRPC. The main factors affecting the timing performance of the MRPC and, accordingly, the optimization scheme are presented.
\end{abstract}


\end{frontmatter}


\section{Introduction}
\begin{figure}[htbp]
\centering
\includegraphics[width=.6\textwidth]{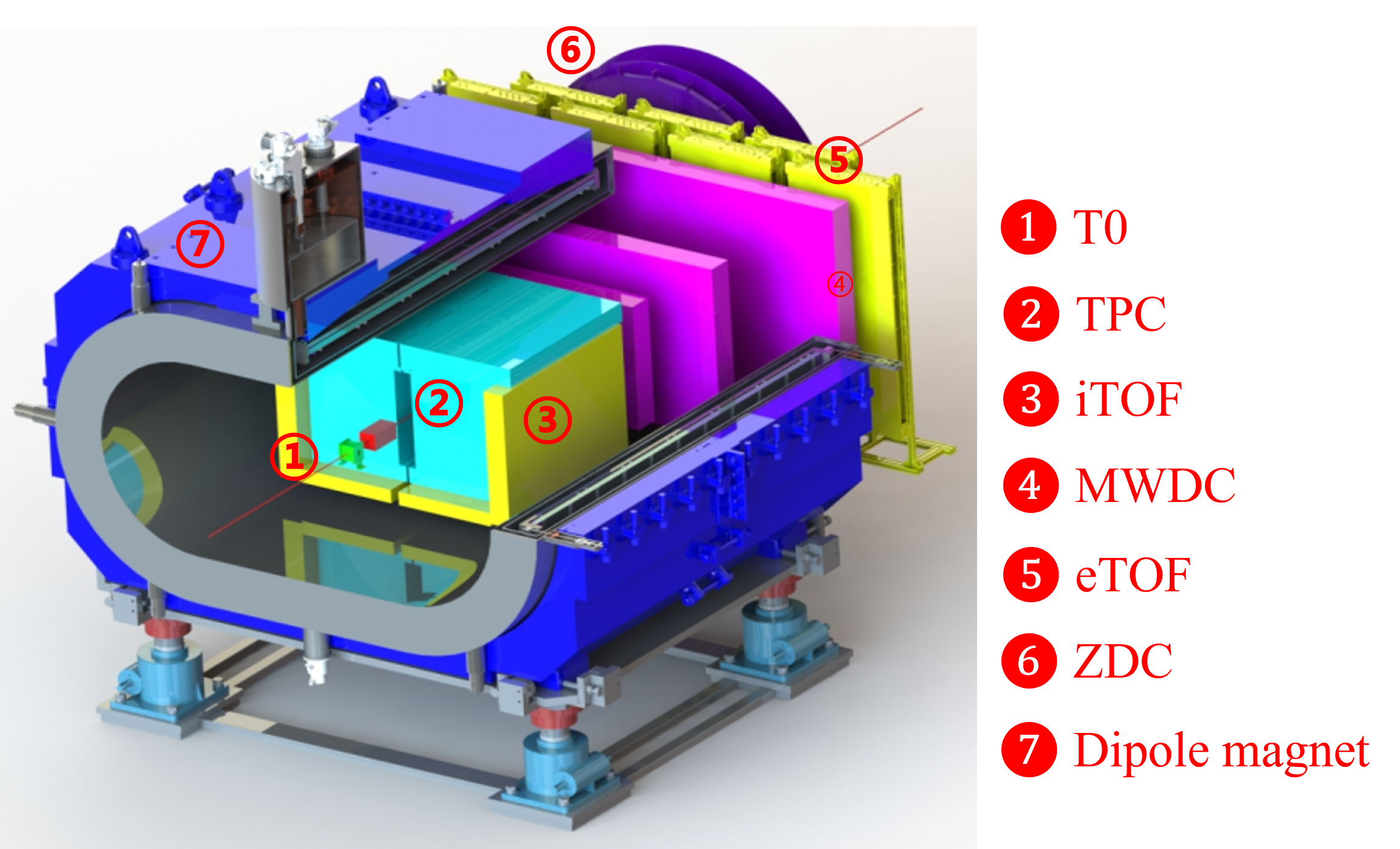}
\caption{Schematic layout of the CEE spectrometer. }
\label{fig:1}
\end{figure}

To study the nuclear matter under high energy or baryon density, and to gain a deep understanding of quantum chromodynamic (QCD), various heavy ion experiments are planned or operational, such as LHC-ALICE~\cite{aamodt2008alice}, RHIC-STAR~\cite{ackermann2003star}, FAIR-CBM~\cite{friese2006cbm}, and NICA-MPD~\cite{toneev2007nica}. The HIRFL-CSR~\cite{xia2002heavy} and high-intensity heavy-ion accelerator facility (HIAF)~\cite{yang2013high} currently under construction can provide beams of different nuclei types in the GeV energy region, which are ideal platforms for studying the properties of dense nuclear matter. A multipurpose spectrometer, the CSR  external-target experiment (CEE), is proposed to run on HIRFL-CSR. The CEE is the first large-scale nuclear experimental device operating in the GeV energy region in China. It adopts fixed-target-mode heavy-ion collisions and will continue its operation with possible upgrades at HIAF\cite{lu2017conceptual}. The conceptual design of a CEE is shown in Fig~\ref{fig:1}.
The CEE spectrometer comprises a set of sub-detectors, including a beam monitor, T0 detector~\cite{hu2017t0}, time projection chamber (TPC)~\cite{li2016simulation}, internal TOF (iTOF ), multiwire drift chamber (MWDC)~\cite{sun2018drift}, external TOF (eTOF), zero-degree calorimeter (ZDC)~\cite{zhu2021prototype}, and large superconducting dipole magnet. In the mid-rapidity or polar angle region (25-107$^{\circ}$) in the laboratory frame, the measurement and identification systems for charged particles are iTOF combined with TPC. Based on the successful operation of multi-gap resistive plate chamber (MRPC) in other heavy-ion experiments\cite{star2007study,zeballos1996new,deppner2012glass}, the CEE also adopts MRPC technology to build its TOF system. The performance requirements of the iTOF-MRPC, including the time resolution, granularity, and the number of channels, were obtained by simulation. We developed and tested an iTOF-MRPC prototype that fulfilled these requirements, as reported in a previous paper\cite{wang2022cee}. However, a study of the iTOF-MRPC prototype also revealed some features that need to be addressed. In this study, we introduce our development of a signal transmission model for iTOF-MRPC, considering main factors such as geometric structure, material, signal propagation, and impedance matching. Experimental tests were performed and the results were compared with the simulation results to verify the model. Key parameters that significantly affected the integrity of the signal were identified. The integrity of the signal can be further improved by choosing an appropriate structure and impedance matching.
\section{Design and Test of high time resolution MRPC}
\subsection{iTOF-MRPC prototype}
A MRPC\cite{zeballos1996new} is gas detector well-known for owing to its good time resolution, high detection efficiency, easy large-area production, and low price. The main factors relevant to the time resolution of the MRPC can be factorised as \cite{Gonzaz_2017}:

\begin{equation}
\sigma_{t}=\sqrt{\frac{d_{gap}\lambda}{N_{gap}}}\frac{U}{\left(\alpha-\eta\right)d_{gap}v}
\label{equ:1}
\end{equation}
where $N_{gap}$ and $d_{gap}$ represent the number and width of gas gaps, respectively. $\alpha-\eta$ is the effective Townsend coefficient, $\lambda$ is the mean free path of ionisation, $v$ is the drift velocity, and $U$ is a factor of order one that accounts for the avalanche statistics. However, the two parameters ($d_{gap}$ and $\alpha-\eta$) are correlated. Their product $\left(\alpha-\eta\right)d_{gap}$ is the natural logarithm of the gas gain for an avalanche developed over a given distance, which is limited by the onset of streamers. Therefore, a narrower gas gap leads to a larger maximum ($\alpha-\eta$) $d_{gap}$ value, and consequently, better time resolution\cite{peskov2018resistive}.

\begin{figure}[htbp]
\centering
\includegraphics[width=.8\linewidth]{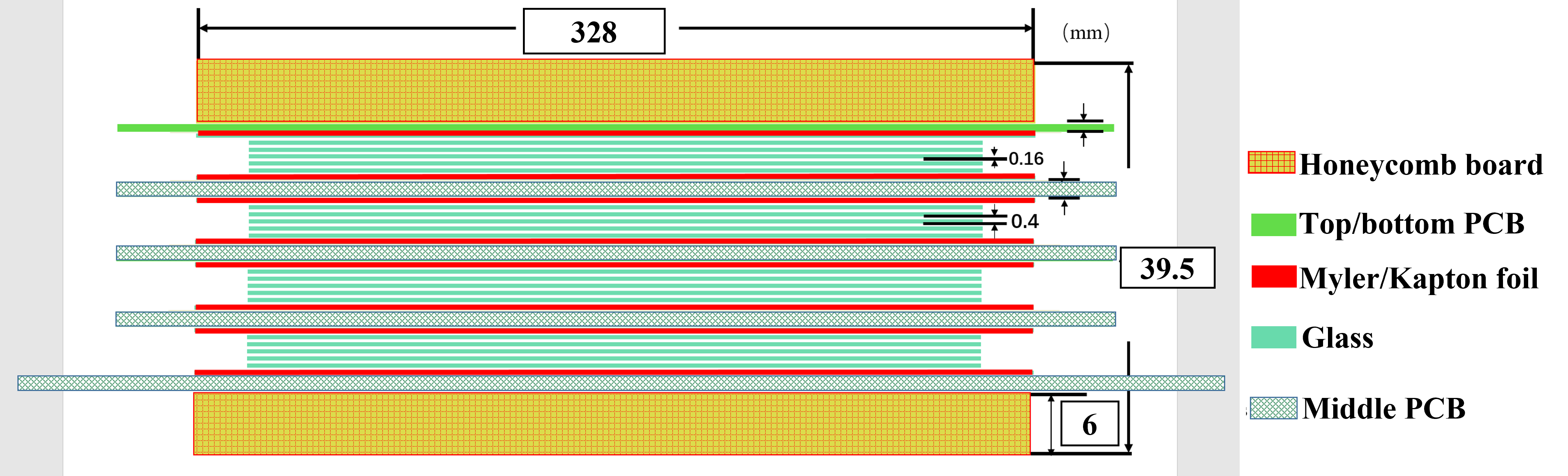}

\caption{Schematic representation of 24-gap iTOF-MRPC prototype, it is divided into four stacks.}
\label{fig:2}
\end{figure}

According to equation \ref{equ:1}, the basic idea of improving the time resolution of the MRPC is to reduce the gap thickness, increase the number of gaps, and select a suitable operating gas. C. Williams designed a 24-gaps MRPC with a 0.16 mm gap width and achieved a time resolution of 20 ps with 95\% efficiency by waveform-sampled readout\cite{an200820}. However, the waveform sampling technique is not suitable for CEE experimental conditions, where high-luminosity heavy-ion collisions and high counting rates are anticipated. The iTOF system plans to use NINO-based front-end electronics (FEE), with optimisation, and an FPGA-based TDC to fit the operation conditions at CEE. The iTOF-MRPC prototype was a four-stack MRPC, as shown in Fig\ref{fig:2}. Each stack contained six gas gaps, and the thickness of the gas gap was 0.16 mm.\cite{wang2022cee}. The prototype contained 32 readout strips, which were 7 mm wide with a 3 mm interval. High voltages were applied to the outermost plates in each stack through graphite electrodes, creating a uniform electric field in the gas gaps, and the glass plates were supported by fishing lines. When a charged particle crosses the detector, it causes primary ionisation in the gas gap, generating an avalanche. The fast-moving electrons in the avalanche induces a fast signal on the readout electrode\cite{Cerron-Zeballos:293074}.

The timing precision of the NINO-based FEE and HPTDC digitisation, commonly used in past heavy-ion experiments\cite{anghinolfi2004nino,llope2012multigap}, was approximately 25 ps. After optimisation, the electronic readout systems of the NINO FEE and FPGA TDC developed for CEE-iTOF have achieved a time jitter of less than 10 ps and a non-uniformity of approximately 2.4\%\cite{9468677}. 

\subsection{Cosmic ray test system}
A cosmic ray test stand system was used to evaluate the performance of the iTOF-MRPC prototype. Two identical MRPC prototypes were assembled and used as time references to each other. The performances of the two MRPCs can be considered identical. According to this hypothesis, we have

\begin{equation}
\sigma_{MRPC}=\frac{\sigma(T_{mrpc1}-T_{mrpc2})}{\sqrt{2}}
\label{equation:MRPC}
\end{equation}
 where $T_{mrpc1}$ and $T_{mrpc2}$ are the measured times when the cosmic ray hits prototypes MRPC1 and MRPC2, respectively. 
The working gases were 90\% Freon-134a, 5\% SF6, and 5\% isobutene. Two plastic scintillator detector readouts by photomultipliers (PMT) were used as the cosmic ray trigger. 

\subsection{Test results}
\begin{figure}[htbp]
\centering
\includegraphics[width=.65\textwidth,height=.55\textwidth]{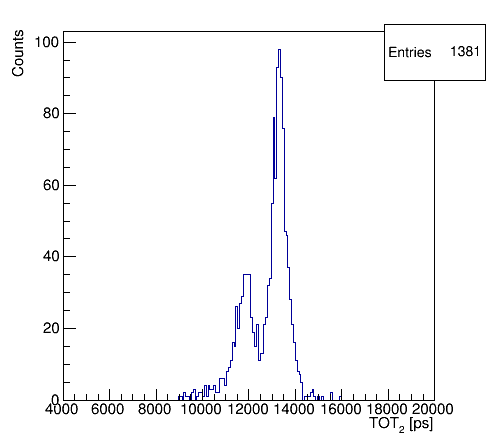}
\caption{TOT distribution measured in the cosmic ray test.The TOT distribution exhibited two distinct peaks that indicates the impedance matching and integrity of the MRPC signal were not perfect.}
\label{fig:5}
\end{figure}

In our lab, the cosmic ray test system was operated smoothly for several months. For both MRPC prototypes, the electronics functioned well. The iTOF-MRPC prototypes achieved an efficiency of more than 95\% and a time resolution of approximately 30 ps at high voltages of $\pm$ 6200V(129 kV/cm)\cite{wang2022cee}. These results fulfilled the requirements of CEE-iTOF. However, there are still flaws in this prototype. The signal width (time-over-threshold, TOT) distribution exhibited two distinct peaks, as shown in Fig\ref{fig:5}, which were caused by multiple reflections of the signal. This indicates that the impedance matching and integrity of the MRPC signal were not perfect. This motivates our further study of signal transmission in the MRPC prototype.

\section{Signal transmission}
\subsection{Raw signal of iTOF-MRPC prototype}
\begin{figure}[htbp]
\centering
\includegraphics[width=.65\textwidth]{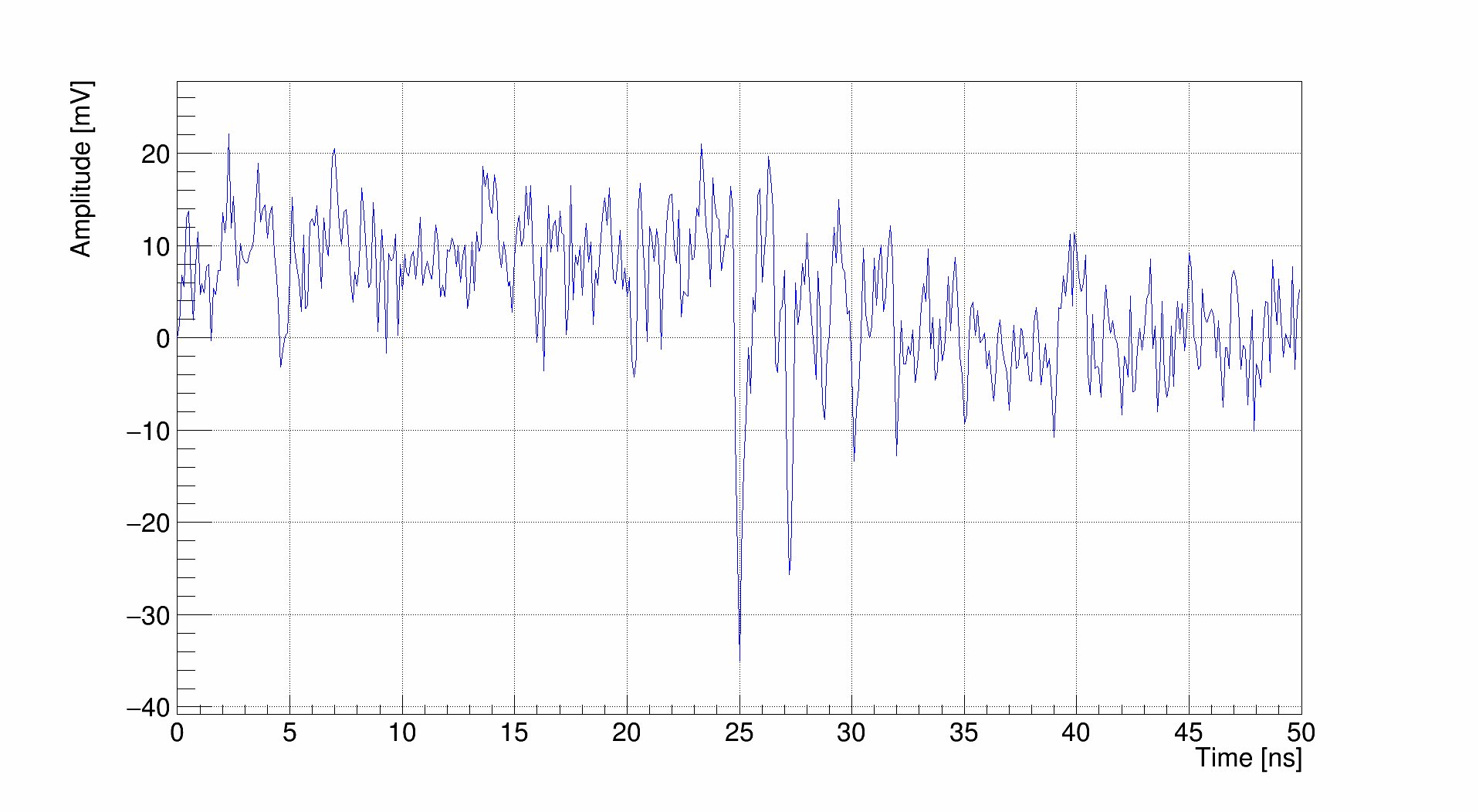}
\caption{Raw signal of iTOF-MRPC prototype collected by high bandwidth oscilloscope.}
\label{fig:7}
\end{figure}

Using a high-bandwidth oscilloscope(4 GHz, 10 Gs/s), we measured raw signal of the iTOF-MRPC prototype, as shown in Fig.\ref{fig:7}. Evidently, multiple reflections exist after the main signal. In the meanwhile, the leading edge of the first peak appears unaffected, and the average rise time is approximately 275 ps.

\subsection{iTOF-MRPC impedance}
\begin{figure}[htbp]
\begin{minipage}[t]{.49\linewidth}
\centering
\includegraphics*[width=.89\textwidth]{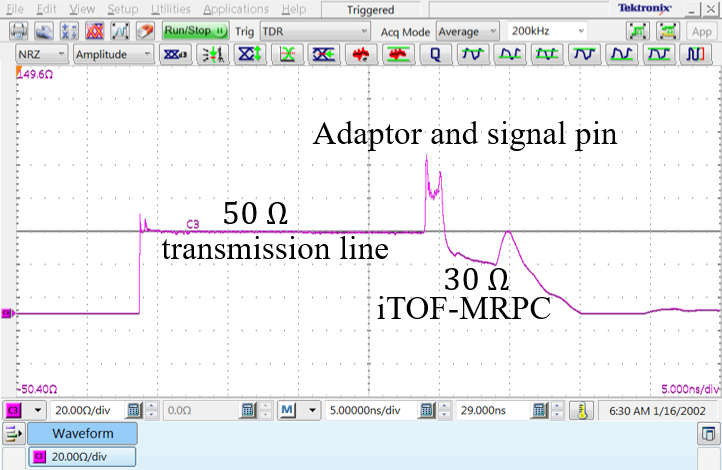}
\centerline{(a)}
\end{minipage}
\hfill
\begin{minipage}[t]{.49\linewidth}
\centering
\includegraphics*[width=.89\textwidth]{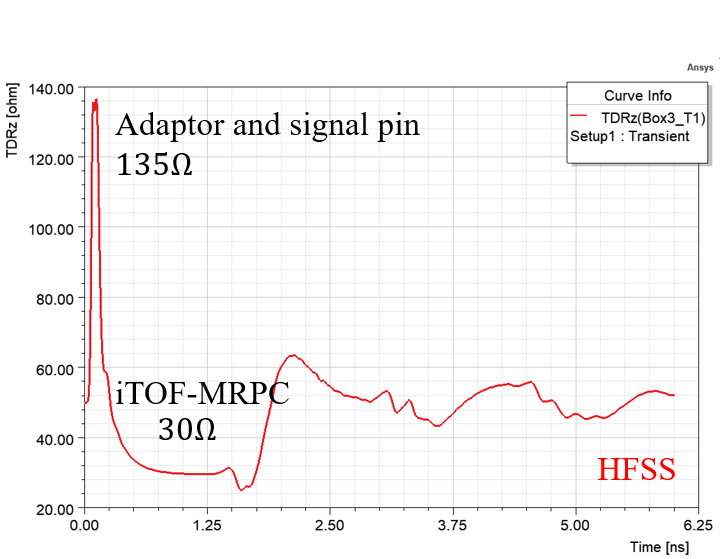}
\centerline{(b)}
\end{minipage}
\caption{(a) MRPC impedance measured using TDR and (b) MRPC impedance results simulated by ANSYS HFSS.}
\label{fig:9}
\end{figure}

A mismatch in the detector impedance can cause reflections. We measured the impedance of the iTOF-MRPC prototype using a time-domain reflectometer (TDR). The characteristic impedance of the transmission line (to the FEE) is 50 $\Omega$, and the impedance of the MRPC strip is 30 $\Omega$, as shown in Fig.\ref{fig:9}. We compared this measurement with the simulation using the ANSYS HFSS software\cite{ANSYS}. A model that was as close as possible to the real iTOF-MRPC prototype was built to calculate the impedance. The results of the simulation and experimental tests are consistent (Fig.\ref{fig:9}), indicating impedance discontinuities between the MRPC strip and the transmission line.
\subsection{Signal transmission model}
\begin{figure}[htbp]
\centering
\includegraphics[width=.55\textwidth]{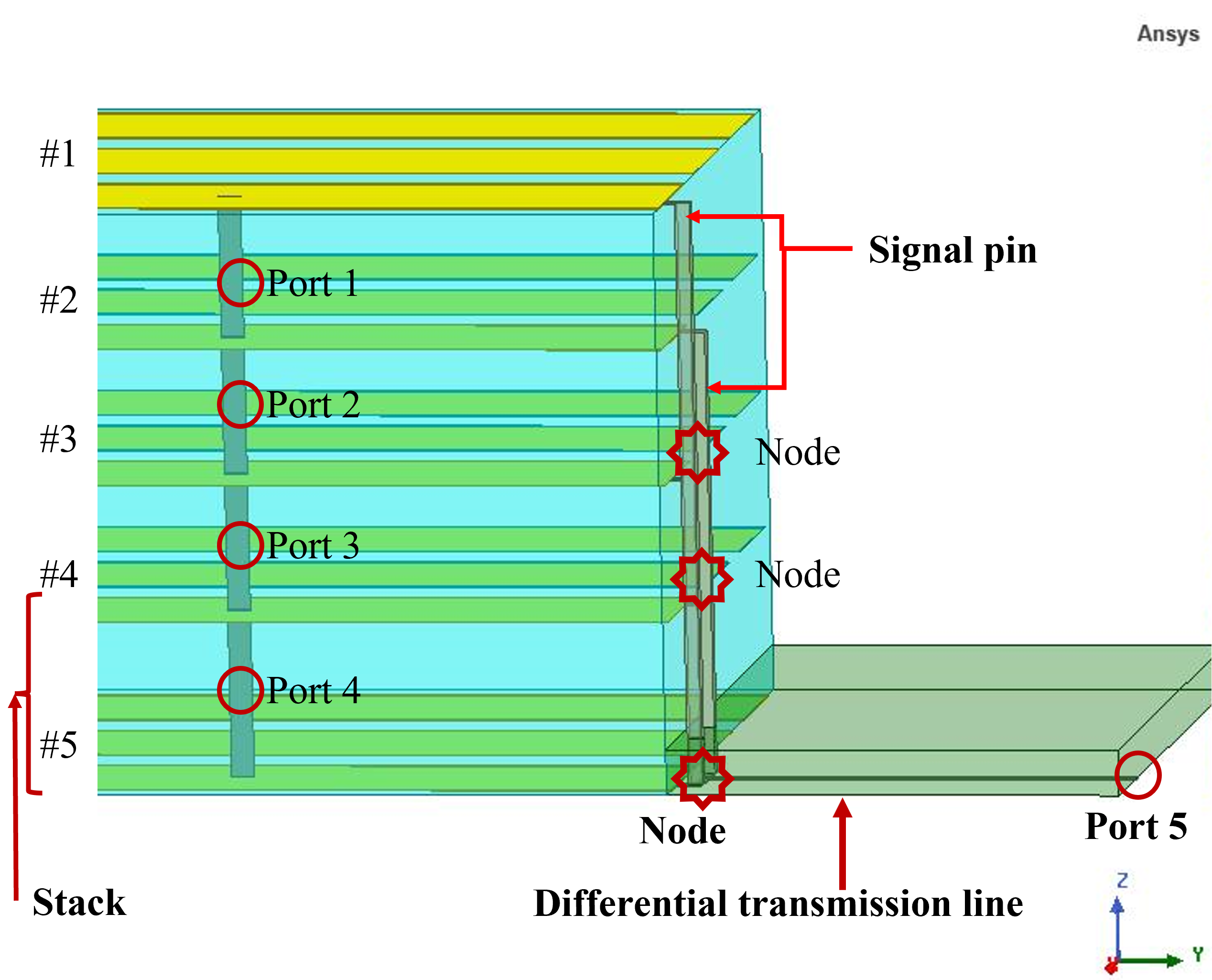}
\caption{The typical iTOF-MRPC model established by HFSS, the junction of the PCB readout strip and the signal pin is the node, and the signal input port and readout port are ports.}
\label{fig:10a}
\end{figure}

A typical iTOF-MRPC model is constructed, as shown in Fig.\ref{fig:10a}. This model simplified the iTOF-MRPC structure by replacing the glass, gas gap, and PCB with an equivalent uniform medium. The permittivity of the equivalent dielectric was adjusted such that the impedance of the model matched the measured value (30 $\Omega$) of the prototype. To speed up the calculation, only 25 strips (five PCBs with five readout strips per PCB) were considered in the simulation, and the detector length was set to 300 mm. The readout strips from the different PCBs (\#1, \#2, \#3, and \#4) were connected to the bottom PCB (\#5) via metal pins. Two signals with opposite polarities, one aggregated from PCB \#1, \#3 and \#5, and the other from PCB \#2 and \#4, are fed to a pair of differential transmission lines, with a length of 20 mm and a differential impedance of 100 $\Omega$, and are then connected to the readout ports (port5, and port6 (not shown in the figure)) at both ends of the MRPC detector. A 1 M$\Omega$ lumped port was placed at the center of each stack. These four ports input a Gaussian-shaped excitation that simulates the induced signal on the readout strip generated by the avalanches in each stack. The Gaussian signal had an amplitude of 1V and a rise time of 230 ps, which is consistent with the real signal. The excitation signal input to each port is delayed by 20 ps sequentially, from the top to the bottom, to simulate the time difference between the induced signals in each stack in the case of particle crossing. To calculate the signal transmission process, a transient solver of ANSYS HFSS was used.

\subsection{Results of simulation}
\begin{figure}[htbp]
\centering
\includegraphics[width=.55\textwidth]{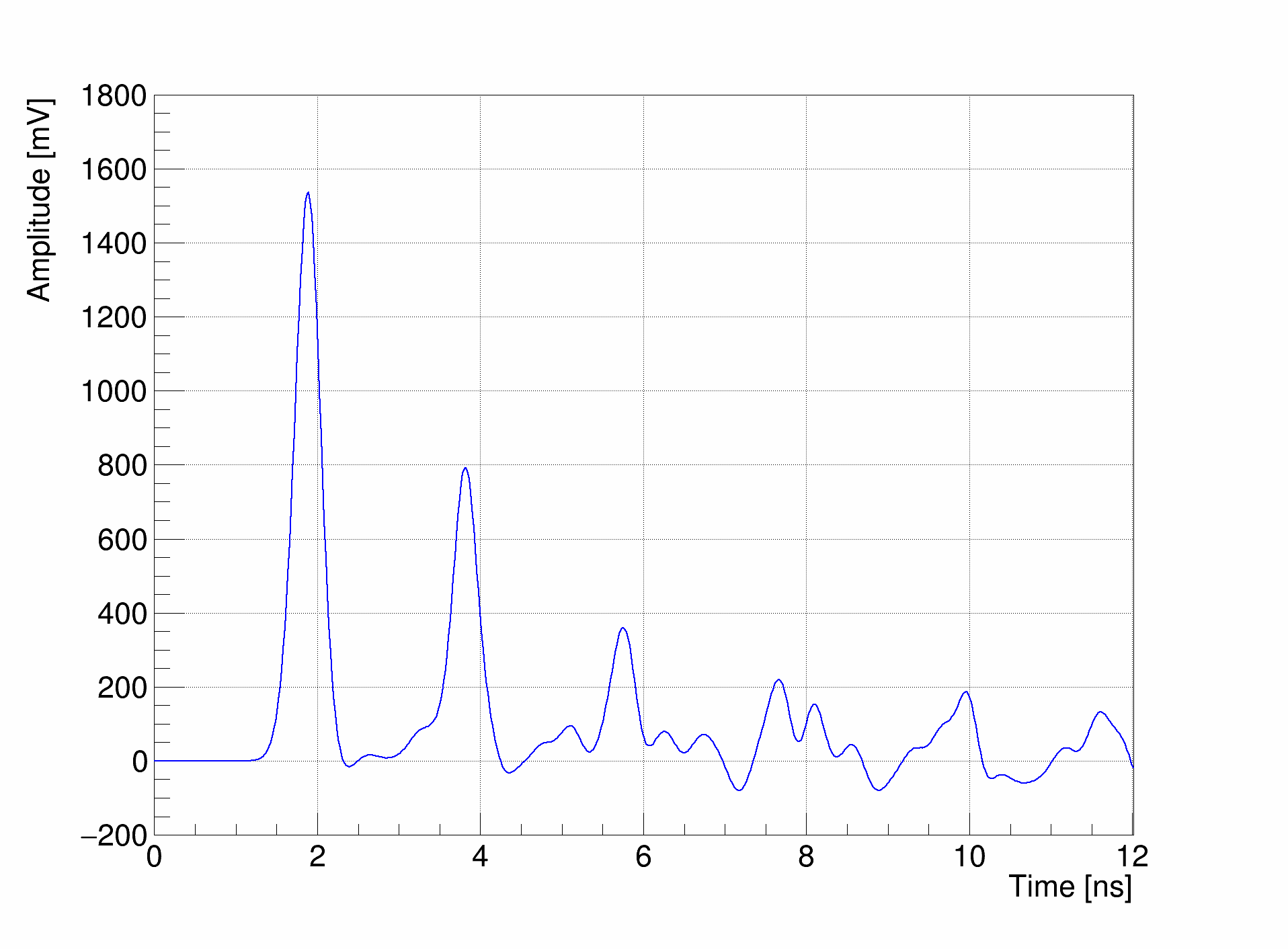}
\caption{The raw signal of iTOF-MRPC simulated by HFSS is consistent with the original signal of iTOF tested by oscilloscope.}
\label{fig:10}
\end{figure}

\begin{figure}[htbp]
\centering
\includegraphics[width=.65\textwidth]{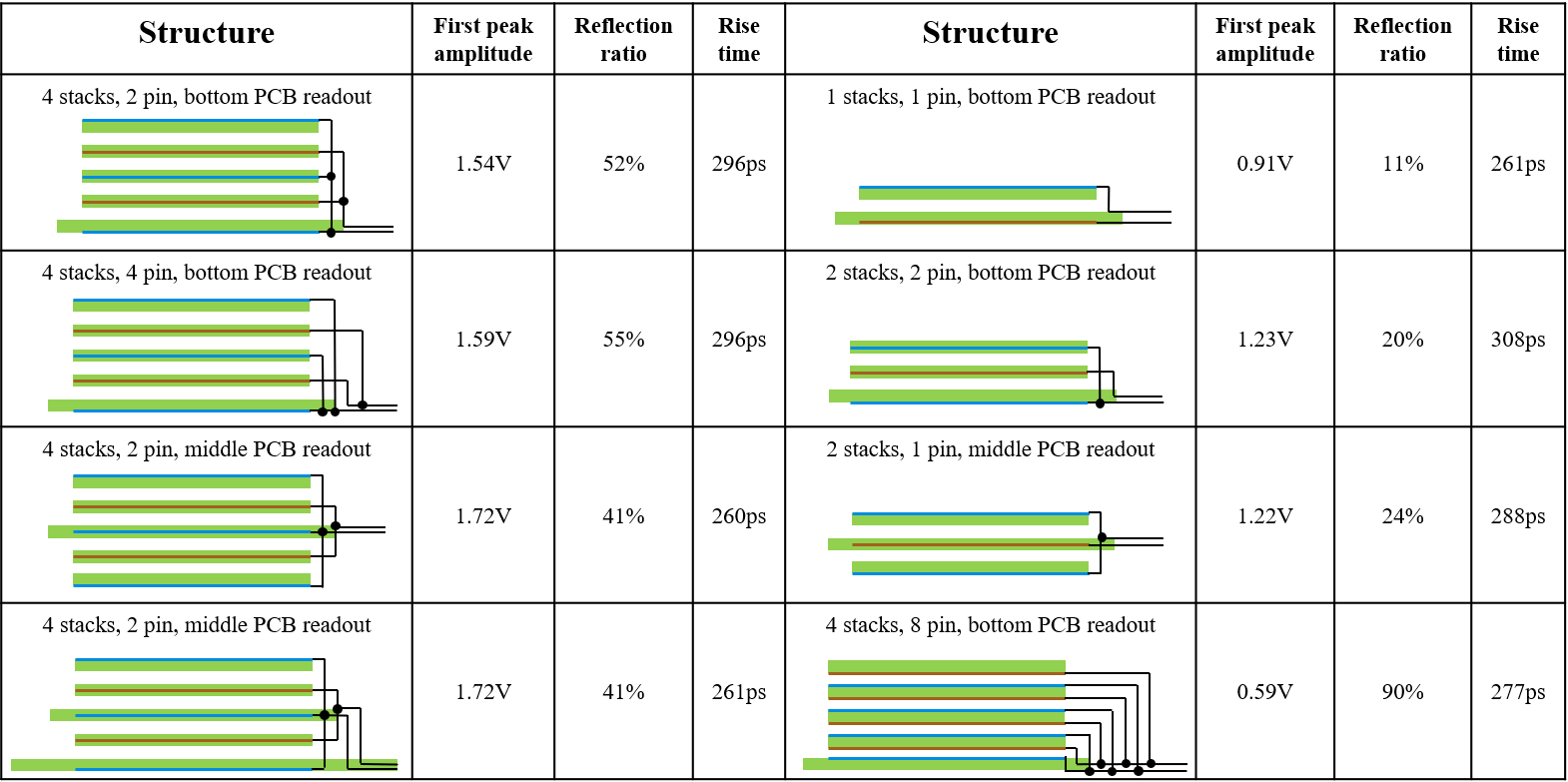}
\caption{Effects of MRPC structures on signal output.}
\label{fig:11}
\end{figure}

The simulation result of the raw signal is shown in Fig\ref{fig:10}, which is consistent with the signal waveform recorded by the oscilloscope(Fig\ref{fig:7}). The reflection is significant and the rise time of the earliest signal increases from 230 ps to 296 ps.
Considering that different structures of the MRPC detector can strongly affect the impedance and transmission, and consequently the output signal, various models are built and simulated to understand how the detector signals of different configurations are propagated. Our strategy is to use three parameters to describe the effect of different structures of MRPC on signal transmission: the amplitude of the first peak, the ratio of amplitude between the second and the first peaks (reflection ratio), and the rise time of the first peak. While pursuing better timing performance of MRPC, faster rise time and lower reflection ratio are favored (higher amplitude of the first peak is also beneficial, but is not the most critical factor). Fig.\ref{fig:11} illustrates a schematic view of the eight models and the corresponding parameters of the simulated results. Only the stack quantity and signal output mode were tuned for these models, while all other settings were kept the same. The simulation results indicate that the greater the number of nodes, the greater is the reflection ratio. When extracting from the middle PCB, the leading edge of the output signal is steeper than that from the bottom PCB. Nevertheless, the rise time of all the signals is larger than that of the input.
To further study the influence of the detector structure on the leading and trailing edges of the signal,the impedance of the iTOF-MRPC strips was maintained at 30 $\Omega$. The effect on the signal shape was studied by adjusting the impedance of the transmission line and the port. 

\begin{figure}[htbp]
\centering
\includegraphics[width=.55\textwidth]{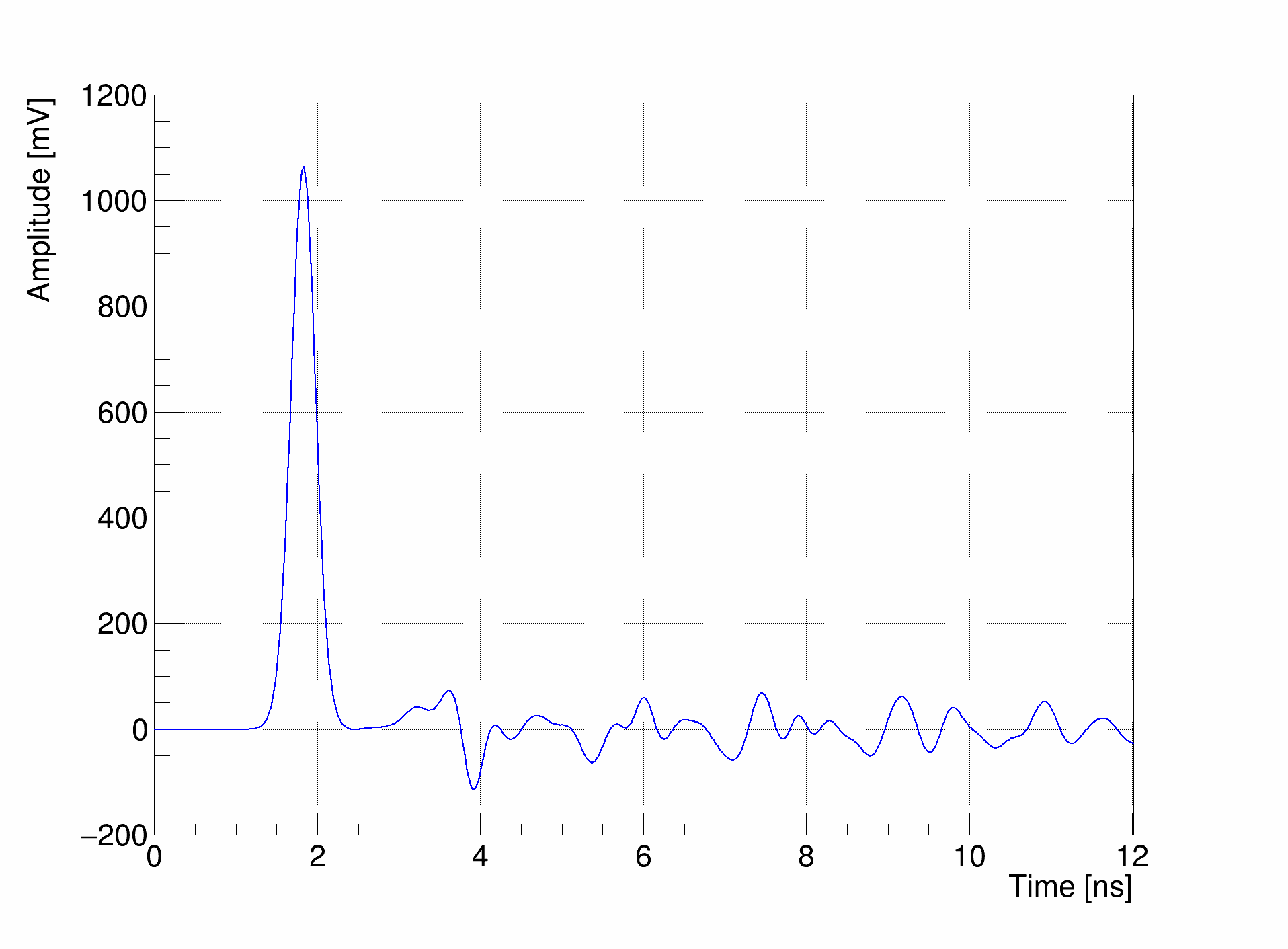}
\caption{Simulated iTOF-MRPC signal readout from middle PCB after impedance matching.}
\label{fig:13}
\end{figure}

The signal integrity is best preserved, and the reflection is minimal when the impedances of the detector, transmission line, and port are the same. However, the structure of the 4 stacks MRPC limits its ability to regulate its own impedance. Therefore, we considered using a resistor to match the impedance of the detector. A 43-$\Omega$ resistor was used in parallel with the 100-$\Omega$ transmission line to match the 30-$\Omega$ iTOF-MRPC readout strip. After impedance matching by the resistor, the signal read out from the middle PCB is shown in Fig.\ref{fig:13}. The simulation finds that the reflection is largely eliminated, and the rise time changes modestly from 261 ps to 266 ps.

Throughout the study, the effect of the impedance of the metal pin was not considered, which was very small owing to its limited length.

\section{Summary}
We designed and built an MRPC prototype for CEE-iTOF, the time resolution of which was found to be approximately 30 ps using cosmic ray tests. To further understand and improve the time resolution of MRPC, ANSYS HFSS was used to simulate the signal transmission process in the MRPC. The simulation results indicate that the number of nodes and readout mode of the MRPC have a significant impact on the performance of the MRPC. The impedance matching between the electronics and detector can be achieved by parallel resistance at the readout terminal, and the impedance of the transmission line has little effect on the detector performance. Further studies,including both simulations and experimental tests, are required.

\section*{Acknowledgments}
The authors thank the high energy physics group of USTC. This project is supported by National Natural Science Foundation of China under grant No. U11927901 and 11975228,the State Key Laboratory of Particle Detection and Electronics under grant No. SKLPDE-ZZ-202202 and the USTC Research Funds of the Double First-Class Initiative under Grant No.WK2030000052.  Special thank goes to Professor Qing Luo from National Synchrotron Radiation Laboratory (China) for help on finite element analysis (ANSYS).

\bibliography{mybibfile}

\end{document}